\documentclass[letterpaper,10pt]{article} 
\usepackage{opticameet3} %% use version 3 for proper copyright statement

\newcommand\authormark[1]{\textsuperscript{#1}}

\usepackage{amsmath,amssymb}
\usepackage[colorlinks=true,bookmarks=false,citecolor=blue,urlcolor=blue]{hyperref} %pdflatex
\begin{document}

\title{0.08 fF, 0.72 nA dark current, 91\% Quantum Efficiency, 38 Gb/s Nano-photodetector on a 45 nm CMOS Silicon-Photonic Platform }

\author{Mingye Fu,\authormark{1, *} and S. J. Ben Yoo\authormark{1, *}}

\address{\authormark{1}Department of Electrical and Computer Engineering, University of California, Davis, CA 95616, USA\\}

\email{\authormark{*}myfu@ucdavis.edu, \authormark{*}sbyoo@ucdavis.edu} %% email address is required

\begin{abstract}
We demonstrated a Germanium-on-Silicon photodetector utilizing an asymmetric-Fabry-Perot resonator with 0.08 fF capacitance. The measurements at 1315.5 nm show 0.72 nA (3.40 nA) dark current, 0.93 A/W (0.96 A/W) responsivity, 36 Gb/s (38 Gb/s) operation at -1V (-2V) bias.
\end{abstract}

\section{Introduction}
Integrated silicon photonics has become a promising technology addressing the ever-increasing demand for high-bandwidth and power-efficient data transmission in high-performance computing\cite{Liang2020} and data center applications\cite{Yoo2022}. Recent advances in integrated photonics showed electronic-photonic integrated circuit (EPIC) optical transceivers being deployed at both ends of optical data links achieved $<$1 pJ/b link energy efficiency\cite{Rizzo2023, Chang2023}. The EPIC optical transceiver demands co-designed, energy-efficient, high-bandwidth photonic components and electronic circuitry integrated with very small parasitics. For instance, one of the most critical photonic components is the Ge photodetector (PD) which is integrated with a transimpedance amplifier (TIA). As the receiver sensitivity is strongly related to the capacitance of the PD through quantum impedance transformation \cite{Miller1989}, detectors with extremely small capacitance ($<$0.1 fF) integrated with zero parasitics can achieve very high sensitivity and link efficiency ($<$0.1 fJ/b at the receiver). A small PD capacitance will also reduce the input capacitance of the TIA which, in turn, can achieve higher bandwidth and lower input noise. In another aspect, a low-capacitance detector is likely to have a small footprint, which leads to a higher integration density of the EPIC transceiver. The small footprint is also helpful in reducing the dark current and consequently improves the optical modulation amplitude (OMA) sensitivity of the receiver. While zero-parasitic integration is available from monolithic CMOS-Silicon Photonic (SiPh) process \cite{Rakowski2020, Baehr-Jones2020}, today's commercial SiPh Ge detectors have typically 15 $\mu m^2$ footprint with 10 fF capacitance including parasitics and 15 nA dark current. It would be desirable to have a nano-photodetector with extremely small capacitance ($<$0.1 fF), low dark current ($<$1 nA), and a small footprint for the next-generation energy-efficient DWDM photonic transceivers.

\begin{figure}[bp]
  \centering
  \includegraphics[width=15.5cm]{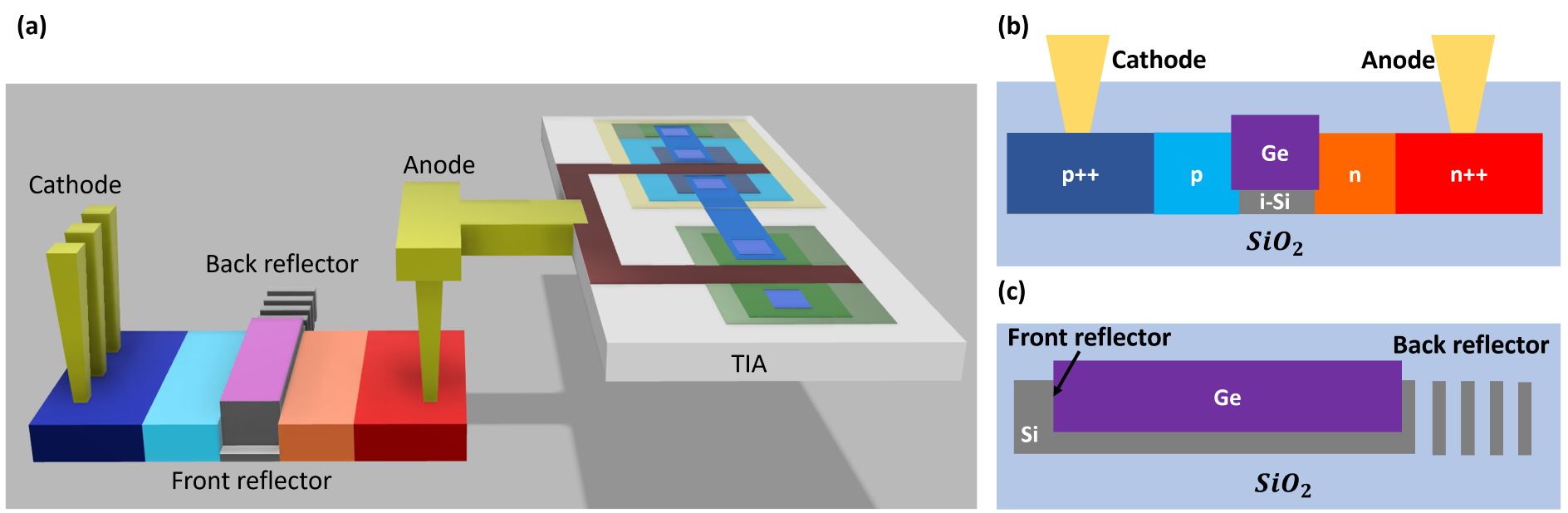}
  \vspace{0.2cm}
\caption{(a) 3D perspective view shows the proposed AFP PD closely integrated with TIA for smallest possible parasitics. (b) AFP PD Cross-section view. (c) Side view showing the resonant cavity.}
\label{PD_design}
\end{figure}

In this paper, we report a Ge-on-Si asymmetric Fabry-Perot (AFP) photodetector with a small Ge footprint of 0.6$\mu m\times$2.2$\mu m$. With an estimated capacitance being extremely low at 0.08 fF, the measured dark current is 0.72 nA under -1 V bias voltage. The quantum efficiency is enhanced by the optimized asymmetric Fabry-Perot resonator achieves a high responsivity of 0.93 A/W at -1 V despite a short cavity length of 2.2$\mu m$. The proposed AFP photodetector is fabricated on GlobalFoundries' 45nm CMOS-SiPh platform (45SPCLO) \cite{Rakowski2020}, allowing it to be monolithically integrated with TIA designed for very low input capacitance, as shown in Fig. \ref{PD_design}a.

\begin{figure}[tbp]
  \centering
  \includegraphics[width=15.5cm]{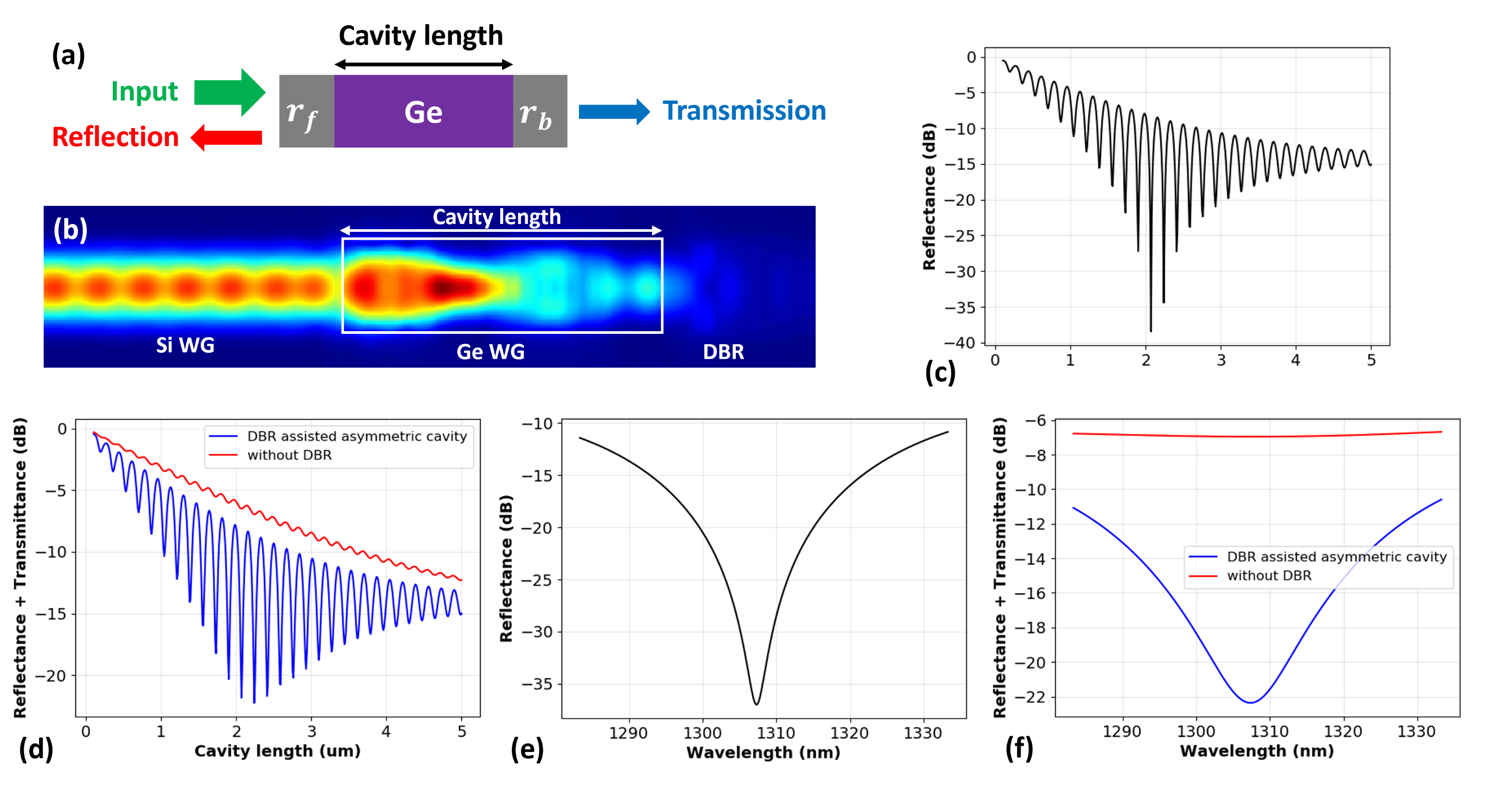}
\caption{(a) Asymmetric Fabre-Perot resonator design. (b) Power profile showing optical coupling and reflection of the AFP PD simulated by FDTD. (c)-(d) Simulation shows the reflection and transmission are nulled at optimized cavity length. Simulation results of (e) reflection spectrum and (f) reflection+transmission spectrum.}
\label{PD_simulation}
\end{figure}

\section{Weakly Resonant Asymmetric-Fabry-Perot photodetector design}
Fig. \ref{PD_design}b and c show the structure of the proposed AFP PD incorporating a waveguide-coupled Ge-on-Si photodetector integrated with an SOI waveguide at the front achieving $>90$\% mode-matched coupling and a distributed-Bragg reflector (DBR) at the back. The front reflector is structured by the SOI-Ge waveguide interface that has $<$10\% reflectivity, which can be fine-tuned by the waveguide geometry. The back reflector has a high reflectivity of $>$95\%. Simulation shows that with optimized cavity design, the AFP PD can almost null the reflection and transmission (defined in Fig \ref{PD_simulation}a) of the AFP resonator, resulting in nearly 100\% absorption of the Ge detector. This is achieved by optimizing the round-trip absorption of the cavity as well as the round-trip phase-matching condition. Fig \ref{PD_simulation}b shows the power profile in the AFP cavity simulated by FDTD. Fig \ref{PD_simulation}c shows that the reflection is nulled near 2.1 $\mu m$ cavity length at 1310 nm, with on-resonance reflectance $<$-30 dB. Maximizing the total power absorbed in the cavity is equivalent to minimizing the reflectance+transmittance (R+T). Fig \ref{PD_simulation}d shows that the R+T minimum ($<$-22 dB) is achieved at 2.2 $\mu m$ cavity length. Fig \ref{PD_simulation}e and f show that the reflection is nulled ($<$ -35 dB) at the center wavelength. R+T is $<$2\% (-17 dB) in 20 nm bandwidth across the center wavelength. It can also be seen from Fig. \ref{PD_simulation} d and f that the DBR-assisted asymmetric cavity improved the total absorption from 75\% to almost 100\% at the center resonance wavelength. The calculations are based on the structure shown in Fig. \ref{PD_design} with Ge waveguide epitaxially grown by selective-area growth. The capacitance is estimated to be 0.08 fF with 0.6 $\mu m$ wide Ge. Note that after fabrication, it is likely that the center wavelength shifts slightly compared to the simulated value, as it is highly sensitive to the detector dimensions.
\section{Characterizations}
\begin{figure}[t]
  \centering
  \includegraphics[width=13cm]{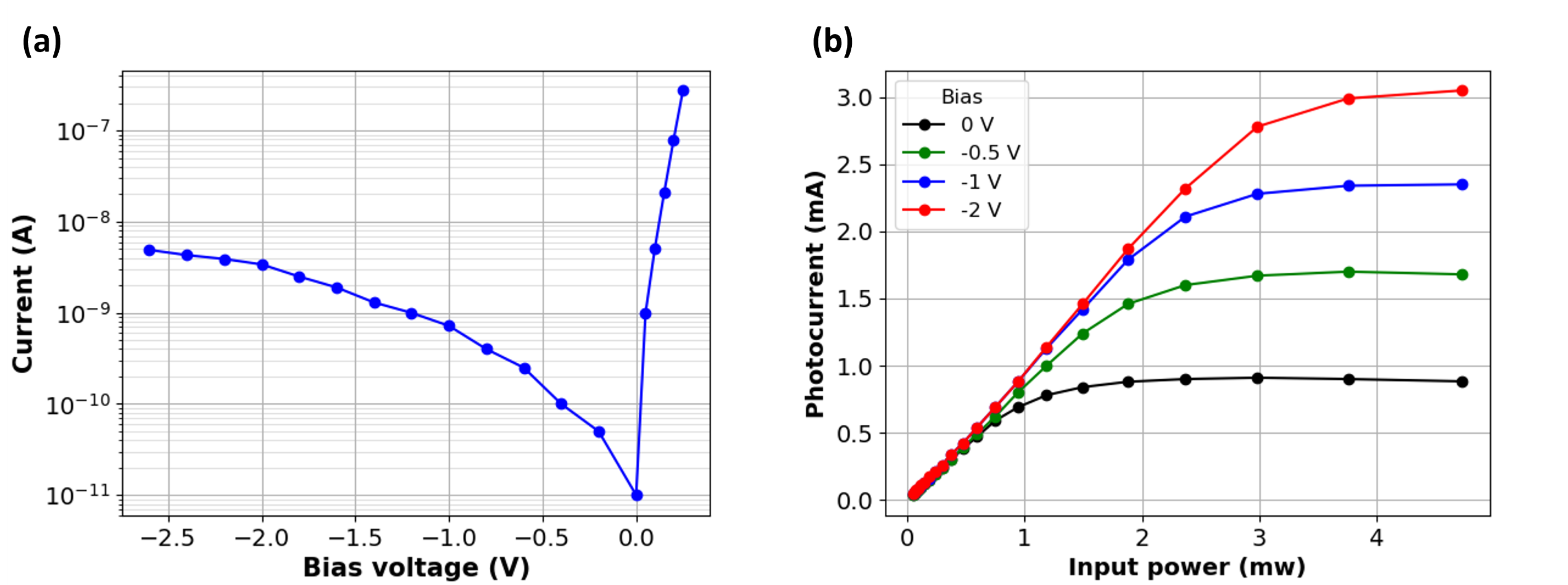}
\caption{(a) Dark current measurement. (b) Photo-current variation at different reverse bias voltages.}
\label{result_DC}
\end{figure}

\begin{figure}[ht]
  \centering
  \includegraphics[width=15.5cm]{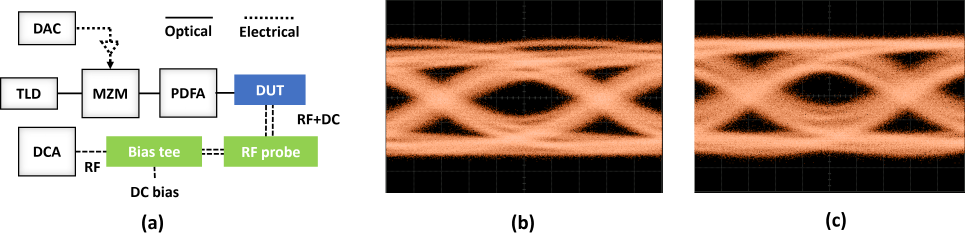}
\caption{(a) High-speed experimental setup. TLD: tunable laser diode; DAC: digital to analog converter; MZM: Mach
Zehnder Modulator; PDFA: Praseodymium-doped fiber amplifier; DUT: device under test; DCA: Digital Communication Analyzer. Eye diagrams of the AFP PD operating at (b) 36 Gbps and (c) 38 Gbps}
\label{result_RF}
\end{figure}
DC characterization is performed utilizing a fiber array as optical input to couple the light into a single-mode waveguide through a grating coupler (GC). At the same time, the grating coupler loss is calibrated using a GC-GC loop-back structure. DC electrical probes are used to apply bias voltage. Fig. \ref{result_DC}a shows the dark current measurement. By the virtual of small footprint, the AFP PD has 0.72 nA dark current at -1 V bias, and 3.40 nA at -2 V bias. The center resonance wavelength (1315.5 nm) is found by sweeping the laser wavelength across 1310 nm and maximizing the photocurrent. Fig. \ref{result_DC}b shows the photocurrent response. The input optical power is controlled by a variable optical attenuator (VOA) with an attenuation tick of 1 dB. When the input power is $<$2 mW, the responsivity is constant for -1 V or -2 V bias. The measured responsivity is 0.93 A/W at -1 V and 0.96 A/W at -2 V. Fig. \ref{result_DC}b also shows the PD saturation under different bias voltage when the input power is high.

The High-speed RF characterization has been carried out utilizing the setup in Fig. \ref{result_RF}a. The DC bias and RF signal are coupled through a bias tee to a high-speed G-S probe. The probe applies DC bias to the PD and at the same time, receives the RF output. The AFP PD is tested using a PRBS-7 pattern at various speeds. A clear and open eye diagram (Fig. \ref{result_RF}c) is observed at 38 Gbps under -2 V bias without any signal amplifier.  It is worth noting that the bias tee (Picosecond 5541A) used in the experiment has a frequency upper limit of 26 GHz, which may be the likely bottleneck of the instrument-limited 38 Gbps operation. Additional measurements with higher-speed instrumentation and measurements including the on-chip CMOS TIAs already integrated with the AFP PDs are in progress.

\section{Conclusions}
Photodetectors with the smallest possible capacitance monolithically integrated with TIA can lead to photonic receivers with high bandwidth and ultra-high sensitivity. In this paper, we proposed and experimentally demonstrated a weakly resonant 1315.5 nm asymmetric-Fabry-Perot photodetector with extremely small capacitance estimated at 0.08 fF, achieving 0.72 nA (3.40 nA) dark current, 0.93 A/W (0.96 A/W) responsivity, 2 mW (3 mW) saturation optical power at -1V (-2V) bias. High-speed data experiment indicates the instrument-limited 38 GHz detector operation. 

\section{Acknowledgement}
This work is supported in part by the Office of the Director of National Intelligence, Intelligence Advanced Research Projects Activity under Award \# 2021-21090200004.


\begin{thebibliography}{99} %% use BibTeX or add references manually

% \bibliography{Mendeley_lib}

\bibitem{Liang2020} D. Liang, G. Kurczveil, Z. Huang, B. Wang, A. Descos, \emph{et al.}, in \emph{Opt. Fiber Commun. Conf. Expo. (OFC)} (2020)

\bibitem{Yoo2022} S. J. Ben Yoo, in \emph{Journal of Lightwave Technology}, 40(8), 2214–2243 (2022)

\bibitem{Rizzo2023} A. Rizzo, S. Daudlin, \emph{et al.}, in \emph{IEEE Journal of Selected Topics in Quantum Electronics}, 29(1), pp. 1-20, (2023)

\bibitem{Chang2023} P. -H. Chang, A. Samanta, P. Yan, M. Fu, \emph{et al.} in \emph{Journal of Lightwave Technology}, 41(21) 6741-6755, (2023)

\bibitem{Miller1989} D. A. B. Miller, in \emph{Opt. Lett.}, 14, 146-148 (1989)

\bibitem{Rakowski2020} M. Rakowski, C. Meagher, K. Nummy, A. Aboketaf, \emph{et al.}, in \emph{Opt. Fiber Commun. Conf. Expo. (OFC)} (2020)

\bibitem{Baehr-Jones2020} T. Baehr-Jones, S. Ardalan, M. Chang, S. Jafarlou, \emph{et al.}, J. Ayala, in \emph{Opt. Express} 31, 24926-24938 (2023)

\end{thebibliography}
\end{document}